\documentstyle [prl,aps,epsfig,floats]{revtex}

\makeatletter           
\@floatstrue
\def\figure{\let\@capwidth\columnwidth\@float{figure}}
\let\endfigure\end@float
\@namedef{figure*}{\let\@capwidth\textwidth\@dblfloat{figure}}
\@namedef{endfigure*}{\end@dblfloat}
\makeatother

\newcommand{\prevl}[3]{Phys.\ Rev.\ Lett.\ {\bf #1}, #2 (#3)}
\newcommand{\prevc}[3]{Phys.\ Rev.\ C {\bf #1}, #2 (#3)}
\newcommand{\zpc}[3]{Z.~Phys.~C {\bf #1}, #2 (#3)}
\newcommand{\plb}[3]{Phys.~Lett.~B {\bf #1}, #2 (#3)}

\newcommand{\be}[1]{\begin{equation}\label{#1}}
\newcommand{\ee}{\end{equation}}
\newcommand{\bI}[1]{{\rm I}_{#1}\!}
\newcommand{\bK}[1]{{\rm K}_{#1}\!}
\def\tf{t_{\rm f}}
\def\vx{\bbox{x}}
\def\vp{\bbox{p}}
\def\vv{\bbox{v}}
\def\la{\left \langle}
\def\ra{\right \rangle}
\def\ch{\cosh}
\def\sh{\sinh}
\def\mt{m_\perp}
\def\pt{p_\perp}

\def\fav{\langle f\rangle(\vp)}

\begin{document}

\def\OurTitlePage
{
\title{Flow effects on the freeze-out phase-space density in heavy-ion 
collisions}

\author{Boris Tom\'a\v sik}
\address{%
Department of Physics, 
University of Virginia, 
P.O. Box 400714,
Charlottesville, Virginia 22904-4714, \\ and
CERN, Theory Division, CH-1211 Geneva, Switzerland}

\author{Ulrich Heinz}
\address{%
Department of Physics, 
The Ohio State University, 
174 West 18th Avenue,
Columbus, Ohio 43210}

\date{November 6, 2001}

\maketitle

\begin{abstract}%
The strong longitudinal expansion of the reaction zone formed in 
relativistic heavy-ion collisions is found to significantly reduce
the spatially averaged pion phase-space density, compared to naive 
estimates based on thermal distributions. This has important 
implications for data interpretation and leads to larger values 
for the extracted pion chemical potential at kinetic freeze-out.
\end{abstract}
}

\draft
\ifpreprintsty
  \OurTitlePage
  \newpage
\else
  \twocolumn[\hsize\textwidth\columnwidth\hsize\csname
    @twocolumnfalse\endcsname
  \OurTitlePage
  \vskip2pc]
\fi

The phase-space density of mesons produced in ultrarelativistic heavy ion 
collisions is an interesting quantity. Its spatial average at freeze-out
(symbolised by $t_{\rm f}$),
 \be{E:aver}
   \fav = {\int d^3x\, f^2(t{>}t_{\rm f},\vx,\vp) \over 
           \int d^3x\, f(t{>}t_{\rm f},\vx,\vp)}\,,
 \ee
can be measured by combining the single-particle momentum spectrum with 
the ``homogeneity volume'' extracted from Bose-Einstein correlation 
measurements \cite{R:bertsch}. Bertsch's original formula \cite{R:bertsch} 
was refined in \cite{R:E877} to exclude contributions from long-lived 
resonances decaying far outside the collision fireball, and a 
relativistically covariant derivation was given in 
\cite{R:phd,R:urs-hab}. With these improvements Bertsch's 
formula reads 
 \begin{equation}
 \label{1}
   \fav =  
   {\sqrt{\lambda(\bbox{p})}\ dn/(dy\, \pt d\pt\,d\phi)
    \over 
    (E_p/\pi^{3/2}) R_s(\vp) \sqrt{R^2_o(\vp) R^2_l(\vp){-}R^4_{ol}(\vp)}}\,.
 \end{equation}
Here $\lambda$ is the intercept at vanishing relative momentum of the
Bose-Einstein correlation function, and the factor $\sqrt{\lambda(\vp)}$ 
corrects for contributions from decays of long-lived resonances in accord 
with the core-halo model \cite{R:core-halo}. The 
numerator is the (thus corrected) invariant momentum spectrum, and the 
denominator contains the homogeneity volume calculated from the Hanbury 
Brown-Twiss (HBT) radii extracted from the Bose-Einstein correlation 
function \cite{R:urs-hab}.

The strongest motivation for measuring the phase-space density of pions at
{\em freeze-out} (i.e. on the last-scat\-tering surface) comes from the
search for multi-boson symmetrization effects \cite{R:pratt}. Analyses
of the hadron yields, spectra, and two-particle correlations from
Pb+Pb collisions at the CERN SPS \cite{R:NN2000} and Au+Au collisions 
at RHIC \cite{R:QM2001} indicate that freeze-out occurs in two 
stages: chemical freeze-out, where the particle yields decouple, 
happens first, reflecting a temperature of around 170 MeV, while
kinetic or thermal freeze-out takes place much later, at temperatures 
around 100--120 MeV. Between these two points the system cools 
adiabatically at constant particle numbers which requires the build-up 
of positive chemical potentials \cite{Bebie:1992ij}. It has been 
speculated that for pions, with only 139 MeV rest mass, this chemical 
potential could approach the Bose condensation limit $\mu_\pi{\,=\,}m_\pi$.
If this were true, measurements of the above type should exhibit large 
values of $\fav$ at small $\vp$. Large phase-space densities also 
generate significant multi-boson symmetrization effects on the pion 
spectra \cite{R:ledn} and Bose-Einstein correlation functions (see 
\cite{R:hsz} and references therein), which lead to a reduction of 
the homogeneity volume extracted from the standard HBT correlation 
analysis \cite{R:urs-hab}. This should cause an additional enhancement of 
$\fav$ as determined by Bertsch's formula (\ref{1}), rendering it at 
the same time unreliable as an estimator of the real average 
phase-space density.
  
Previous analyses of heavy-ion data at the AGS \cite{R:E877} and SPS 
\cite{R:ferenc,R:phd} concluded that there was no evidence for un\-usually 
large pion freeze-out phase-space densities at these collision 
energies. This conclusion was based on a comparison of $\fav$ from 
Eq.~(\ref{1}) with a Bose-Einstein distribution, yielding rough agreement 
when inserting a kinetic freeze-out temperature of around 100--120 MeV (as 
extracted by other methods \cite{R:NN2000}) and a small or vanishing 
pion chemical potential. In some cases \cite{R:phd,R:cramer} comparison 
was made with a transversely boosted Bose-Einstein distribution, in order 
to account for transverse flow effects. We here point out that all these 
comparisons may have been misleading since they neglected a strong reduction 
effect on the spatial averaging originating from the {\em longitudinal} 
expansion of the source. We show that for longitudinally expanding sources 
the spatially averaged phase-space density is smaller than its thermal 
value in the local rest frame, and that accounting for this effect is 
likely to considerably increase the extracted value of the pion chemical 
potential at kinetic freeze-out. In this paper we concentrate on a 
theoretical exposition of the basic mechanism and leave a (re)analysis 
of existing data for later.  

For a qualitative understanding of the effect under discussion, let us 
for simplicity assume a fireball of constant matter density. We fix the 
momentum and focus on the averaging over position space. If the 
(thermalized) fireball expands, different parts moving with different 
velocities contribute to the production of particles with fixed momentum 
$\vp$ with different rates, given by Bose-Einstein distributions {\em boosted} 
by the local flow velocity. The phase-space density at momentum $\vp$  
then becomes a function of position space. For an average of type 
(\ref{E:aver}), where the function is weighted with itself, the result 
is {\em always smaller} than the maximum of the function (here 
corresponding to zero flow velocity). In general the $\vp$-dependence 
of the spatial average will no longer be of Bose-Einstein form.
[Obviously, there is no position dependence of the emission rate in
a {\em non-expanding} source for which the spatial average would thus
coincide with the underlying distribution.]

We illustrate this effect quantitatively within a specific model for the 
expanding source. We will assume a thermalized and longitudinally 
boost-invariant fireball. This simple model is expected to provide a 
good description of pions emitted near midrapidity at RHIC and LHC 
energies. We show that the calculated $\pt$-dependence of the spatially
averaged phase-space density is flatter than a Bose-Einstein distribution 
with the assumed freeze-out temperature. The major contribution to this 
flattening arises from the {\em longitudinal} expansion which strongly 
{\em reduces} $\fav$ relative to the corresponding static Bose-Einstein 
distribution. Transverse flow introduces a weaker additional flattening 
by shifting weight from low to high transverse momenta $\pt$.

The phase-space density $f(t,\vx,\vp)$ is related to the source emission 
function $S(x,\vp)$ \cite{R:urs-hab} by
 \be{E:psd-S}
   f(t,\vx,\vp) = \frac{(2 \pi)^3}{E_p} \, 
   \int_{-\infty}^t dt'\, S(t',\vx + \vv (t'-t),\vp) \, ,
\ee
where $\vv{\,=\,}\vp/E_p$ is the velocity corresponding to the momentum 
$\vp$. The prefactor $(2\pi)^3/E_p$ ensures the standard normalisation
 \begin{eqnarray}
 \label{E:f-norm}
   \bar N = \int {d^3x\,d^3p \over (2\pi)^3}\, f(t,\vx,\vp) 
          = \int {d^3p \over E_p}\,d^4x\, S(x,\vp)\,.
 \end{eqnarray}
We take $\bar N $ as the average number of ``directly emitted'' pions, 
including those from the decays of shortlived resonances but excluding
those from resonances with long lifetimes \cite{R:E877,R:ferenc}. For 
simplicity, we follow Refs. \cite{Kataja:1990tp,Gerber:1990yb} and assume 
that the change in shape and normalisation of the thermal pion distribution 
at freeze-out caused by adding pions from shortlived resonance decays can 
be effectively absorbed by giving the pions a positive chemical potential. 
Although this cannot replace a full resonance decay calculation using the 
proper decay kinematics \cite{Sollfrank:1990qz}, it qualitatively 
reproduces the decay-induced low-$\pt$-enhancement of the measured pion 
spectra \cite{Kataja:1990tp,Gerber:1990yb,Sollfrank:1990qz}.

Eq.~(\ref{E:aver}) is brought into the form (\ref{1}) by following the 
steps outlined in Sec. 3.4 of Ref. \cite{R:urs-hab}. In order to include 
all ``directly emitted'' pions, the time $t$ on the right hand side of
Eq.~(\ref{E:aver}) must be later than the time at which the last
pion was emitted. Due to Liouville's theorem, the average phase-space 
density (\ref{E:aver}) remains constant after completion of the freeze-out 
process. This allows \cite{R:phd} to replace in (\ref{E:aver}) the 
spatial integrals at constant time $t$ by integrals over the {\em freeze-out 
hypersurface} $\Sigma_{\rm f}(x)$ describing the last pion scattering or 
production points:
 \be{E:aver2}
   \fav = {\int_{\Sigma_{\rm f}} p\cdot d^3\sigma(x) \, f^2(x,\vp) \over 
           \int_{\Sigma_{\rm f}} p\cdot d^3\sigma(x) \, f(x,\vp)}\,.
 \ee
If $\Sigma_{\rm f}(x)$ is parametrised by an $\vx$-dependent freeze-out
time $t_{\rm f}(\vx)$, then
 \begin{eqnarray}
   \lefteqn{\int_{\Sigma_{\rm f}} p\cdot d^3\sigma(x)\, f^n(t,\vx,\vp) =} 
   \hspace{4em} &&
 \nonumber \\ && 
   E_p \int d^3x\, (1 - \vv{\cdot}{\nabla\!}_{x\,} t_{\rm f}(\vx))\;
   f^n(t_{\rm f}(\vx),\vx,\vp)\, .
 \label{E:hyper}
 \end{eqnarray}

For the model emission function we take \cite{R:hydromodel}
 \begin{eqnarray}
   S(x,p) = {\mt \ch(y{-}\eta) \over (2\pi)^3}\,
            {\delta(\tau{-}\tau_{\rm f}) \over
             \exp\bigl[{p\cdot u(x){-}\mu(r)\over T}\bigr]{-}1}
 \label{E:model}
 \end{eqnarray}
where
 \[
   p{\cdot}u(x){=}\mt \ch(y{-}\eta) \ch\zeta(r)
                - \pt \cos(\phi{-}\varphi) \sh\zeta(r).
 \]
We use longitudinal proper time $\tau{=}\sqrt{t^2{-}z^2}$, space-time rapidity 
$\eta=$ ${1\over 2}\ln[(t{+}z)/(t{-}z)]$, and transverse polar coordinates 
$(r,\,\varphi)$ to parametrise $x$. The pion momentum is 
$p{\,=\,}(\mt\ch y,\,\pt\cos\phi,\,\pt\sin\phi,\,\mt \sh y)$.
Eq.~(\ref{E:model}) describes a longitudinally infinite source with 
boost-invariant longitudinal expansion and sharp freeze-out at proper 
time $\tau_{\rm f}$. Transverse expansion is parameterised by 
the transverse flow rapidity $\zeta(r)$, to be specified below. 
The $r$-dependent chemical potential $\mu(r)$ allows to discuss different
transverse density profiles of the source at freeze-out. While probably 
too simple for a quantitative comparison with data, this model is 
technically convenient and allows to investigate the effects of 
longitudinal and transverse flow as well as the influence of the 
chemical potential on the average phase-space density $\fav$.

We now evaluate Eq.~(\ref{E:aver2}) for midrapidity pions ($y{\,=\,}0$).
We can rotate the coordinate frame such that $\phi{\,=\,}0$. From 
(\ref{E:psd-S}) we obtain for the phase-space density immediately 
after freeze-out
 \be{E:fx}
   f(\tf(\vx),\vx,\vp) = \left[\exp\left(p\cdot u(x_{\rm f}){-}\mu(r)
                         \over T\right){-}1\right]^{-1}\,,
 \ee
with $x_{\rm f}{\,=\,}(\tf(\vx),\vx)$. For the integration in 
Eq.~(\ref{E:hyper}) we use the freeze-out surface at 
$\tau{\,=\,}\tau_{\rm f}$ with integration measure
 \begin{eqnarray}
   \bigl(1-\vv{\cdot}{\nabla\!}_{x\,}\tf(\vx)\bigr)\,d^3x =   
   \tau_{\rm f}\ch\eta\, d\eta\, r\, dr\, d\varphi \, .
 \end{eqnarray}
To obtain $\fav$ from (\ref{E:aver2}) we need ($E_p{\,=\,}\mt$ for 
$y{\,=\,}0$)
 \begin{eqnarray}
   A_n(\pt) = \mt\tau_{\rm f} \int {\ch\eta\, d\eta\, r \, dr\, d\varphi
   \over \left[\exp \left(\frac{p\cdot u(x_{\rm f}){-}\mu(r)}{T} \right) - 1 
   \right]^n}
 \label{E:adef}
 \end{eqnarray}
for $n{\,=\,}1,2$. Note that $A_1(\pt)$ is, up to a factor $(2\pi)^{-3}$,
the invariant momentum spectrum $E_p (dN/d^3p)$. 

We evaluate $A_n$ by expanding the Bose distribution as
$(e^x{-}1)^{-1} = \sum_{k=1}^{\infty} e^{-k x}$. This reduces (\ref{E:adef})
to a sum of exponential integrals. We now integrate over $\eta$ and 
$\varphi$ in the standard way \cite{Wiedemann:1996au}, obtaining 
Bessel functions $\bK{1}$ and $\bI{0}$, respectively. Only the radial 
integration must be done numerically. With the shorthands
\begin{mathletters}
\label{E:shands}
\begin{eqnarray}
\label{E:alpha}
\alpha_\perp(r)& = & \pt\sh\zeta(r)/T \\
\label{E:beta}
\beta_\perp(r) & = & \mt\ch\zeta(r)/T 
\end{eqnarray}
\end{mathletters}
we find
 \begin{mathletters}
 \label{E:results}
 \begin{eqnarray}
 \label{E:res-main}
    \la f \ra (\pt) &=& \frac{A_2(\pt)}{A_1(\pt)}\, ,
 \\
 \label{E:res-A1}
    A_1(\pt) &=& \sum_{k=1}^{\infty} \tilde A_k(\pt)\,,
 \\
 \label{E:res-A2}
    A_2(\pt) &=& \sum_{k=2}^{\infty} (k{-}1) \tilde A_k(\pt)\,,
 \\
 \label{E:res-A3}
    \tilde A_k(\pt) &=& 4 \pi \mt \tau_{\rm f} \int_0^\infty r\,dr\, 
       e^{k[\mu(r)/T]}
 \nonumber\\
    &&\times \bI{0}\bigl(k\,\alpha_\perp(r)\bigr)\,
       \bK{1}\bigl(k\,\beta_\perp(r)\bigr)\,.
 \end{eqnarray}\end{mathletters}%
We will consider two models for the transverse density profile:
a box-like density
\begin{mathletters}
\label{E:profs}
\be{E:box}
\mbox{[box]} \qquad \mu(r) = 
\left \{ 
\begin{array}{ll}
  \mu_B \quad \mbox{for} \quad & r\le R_{\rm box} \\
  - \infty \quad & \mbox{otherwise}
\end{array} , \right .
\ee
and a Gaussian profile
\be{E:Gauss}
\mbox{[Gauss]} \qquad \mu (r) = \mu_G - T \frac{r^2}{2R^2_{\rm Gauss}}\, .
\ee\end{mathletters}%
Here $\mu_B$ and $\mu_G$ are the chemical potential values {\em in 
the center} of the fireball. For the Gaussian profile, $\mu_G$ will be 
{\em larger} than the global (averaged) chemical potential used in  
chemical analyses of particle abundances \cite{Heinz:1999kb}.

As a warm-up, let us consider the case without transverse flow, 
$\zeta(r){\,=\,}0$, in the Boltzmann approximation (i.e. keeping
only the first term in the sums over $k$). Then
 \be{E:noflow}
   \la f^{\rm Boltz}_{\zeta=0} \ra(\pt) = 
   \bar \lambda_\pi \frac{\bK{1}\left(2\frac{\mt}{T}\right)}
                              {\bK{1}\left(\frac{\mt}{T}\right)}\, ,
 \ee
where $\bar\lambda_\pi$ is the spatially averaged pion fugacity
 \be{E:fug}
   \bar \lambda_\pi = \la e^{\mu/T} \ra \equiv
   \frac{\int r\,dr\, e^{2\mu(r)/T}}
        {\int r\,dr\, e^{\mu(r)/T}} \,,
 \ee
the average being taken with the transverse matter density 
$\rho(r)\sim e^{\mu(r)/T}$. A box-like density gives
$\bar\lambda_\pi{\,=\,}e^{\mu_B/T}$ whereas a Gaussian profile 
leads to $\bar\lambda_\pi{\,=\,}\frac{1}{2} e^{\mu_G/T}$.

\begin{figure}
\vbox{
   \begin {center}
      \epsfig{file=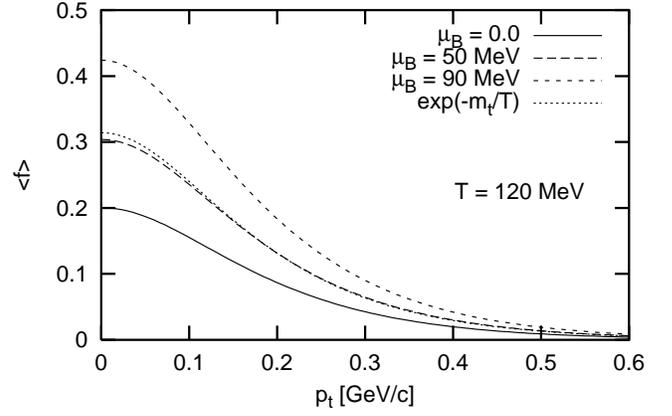,scale=1.7}
   \end {center}
\caption{Average phase-space density $\la f \ra(\pt)$ at rapidity
$y{\,=\,}0$ for pions from a longitudinally expanding fireball with 
Boltzmann-distributed particle momenta, for a temperature of 120~MeV
and three representative values for the pion chemical potential.
\label{F:bolz}
}
}
\end{figure}

In Fig.~\ref{F:bolz} we show the result (\ref{E:noflow}) for 
$T{\,=\,}120$\,MeV and three values for the average pion fugacity
$\bar\lambda_\pi$. 
According to (\ref{E:noflow}) the curves do not depend on the transverse 
density profile. The calculation in Fig.~\ref{F:bolz} was done with a 
box-like density profile for which $\mu_B$ can be considered as the 
global chemical potential. The same curves would be obtained in the 
Gaussian model by adjusting $\mu_G$ such that 
$e^{\mu_G/T}{\,=\,}2\,e^{\mu_B/T}$ (leading to a twice larger fugacity 
$\lambda_\pi(r{=}0){\,=\,}2\bar\lambda_\pi$ in the fireball center).

The dotted line in Fig.~\ref{F:bolz} is shown for reference and represents
a Boltzmann distribution with vanishing pion chemical potential
for the same temperature as used in the calculations  (i.e. the 
original thermal pion momentum distribution in the local rest frame). 
Comparing it with the solid curve one sees that, even without transverse 
flow, at low $\pt$ the spatially averaged phase-space density of the 
longitudinally expanding source is about 35{\%} lower than its value 
in the local rest frame. In order to obtain an $\la f \ra(\pt)$ which 
looks like a Boltzmann distribution with vanishing pion chemical 
potential, we need in fact an average pion fugacity 
$\bar\lambda_\pi{\,\approx\,}1.55$. For practical purposes this 
interpretation is better presented the other way around: if the 
curve corresponding to $\bar\lambda_\pi{\,\approx\,}1.55$ (or 
$\mu_B{\,=\,}50\,\mbox{MeV}$ at $T{\,=\,}120$\,MeV) was measured and 
then fitted by a Boltzmann distribution with $T{\,=\,}120$\,MeV, one 
would erroneously conclude $\mu_B{\,=\,}0$.

For large $\mt$, $\bK{1}(2\mt/T)/\bK{1}(\mt/T)\approx e^{-m_\perp/T}$, 
and this type of error no longer arises. Unfortunately, this requires 
for pions much larger values of $\pt$ than presently experimentally 
accessible. The reason for the strong reduction at low $\pt$, resulting 
from taking the spatial average, is that this average extends over
a large longitudinal homogeneity length, and that for a longitudinally 
expanding source $f(x;y{=}0,\pt)$ agrees with the static Boltzmann 
distribution $\exp(-\mt/T)$ only at the point $\eta{\,=\,}0$, being 
smaller everywhere else by a factor $\exp[-\mt(\ch\eta-1)]$. For large 
$\pt$ the longitudinal homogeneity length decreases \cite{R:urs-hab} 
and the reduction due to spatial averaging becomes less severe.

\begin{figure}[ht]
\vbox{
   \begin {center}
      \epsfig{file=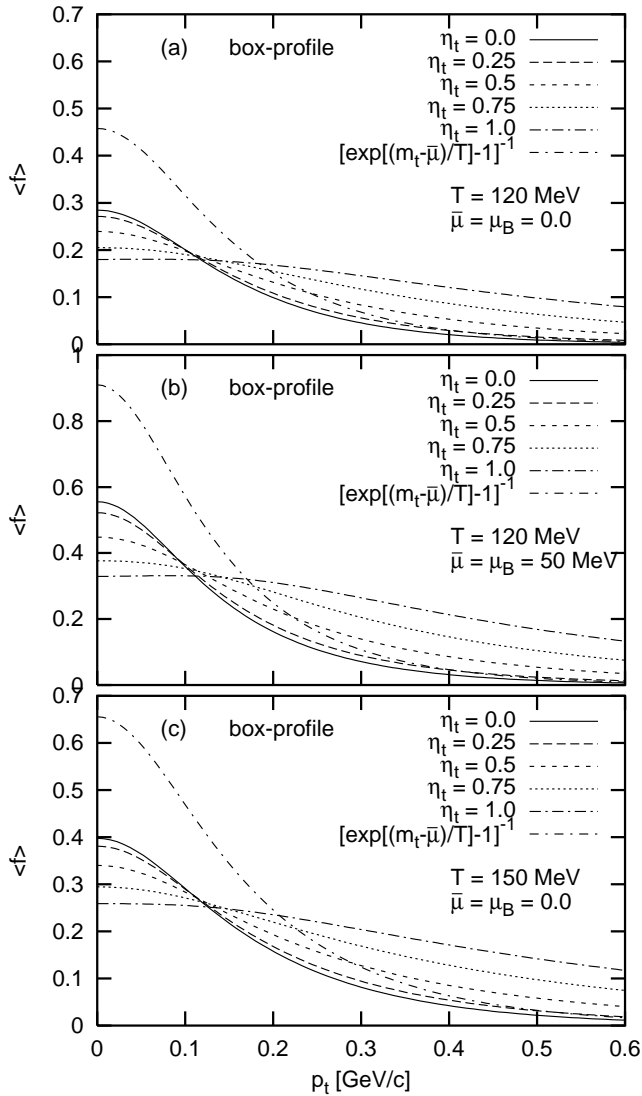,scale=1.7}
   \end {center}
\caption{Average phase-space density $\la f \ra$ for pions with $y{\,=\,}0$ 
from a longitudinally and transversally expanding fireball with Bose-Einstein 
distributed momenta of temperature 120 MeV. The transverse flow is 
controlled by $\eta_t$. The transverse density profile has the shape of
a box.
\label{F:flow}
}
}
\end{figure}

We now discuss the more general case of a source with additional 
transverse flow, including the full Bose-Einstein distribution by 
summing over all $k$ in Eqs.~(\ref{E:results}). The transverse flow 
profile $\zeta(r)$ influences the radial averaging of the factors 
containing the pion chemical potential, and it is no longer possible 
as in (\ref{E:noflow}) to factorize an effective average pion fugacity 
$\bar\lambda_\pi$ from the result for $\la f\ra(\pt)$. For the 
transverse flow rapidity we take a linear profile with slope $\eta_t$,
 \be{E:flowprofile}
   \zeta(r) = \eta_t\, \frac{r}{r_{\rm rms}},
 \ee
where $r_{\rm rms}{\,=\,}R_{\rm box}/\sqrt{2}$ and 
$r_{\rm rms}{\,=\,}\sqrt{2}\,R_{\rm Gauss}$, respective\-ly, are the rms 
radii corresponding to the two different transverse density profiles.
Thus $\eta_t{\,=\,}\zeta_{\rm rms}$ can  be interpreted as the average 
(rms) transverse flow rapidity.

\begin{figure}[th]
\vbox{
   \begin {center}
      \epsfig{file=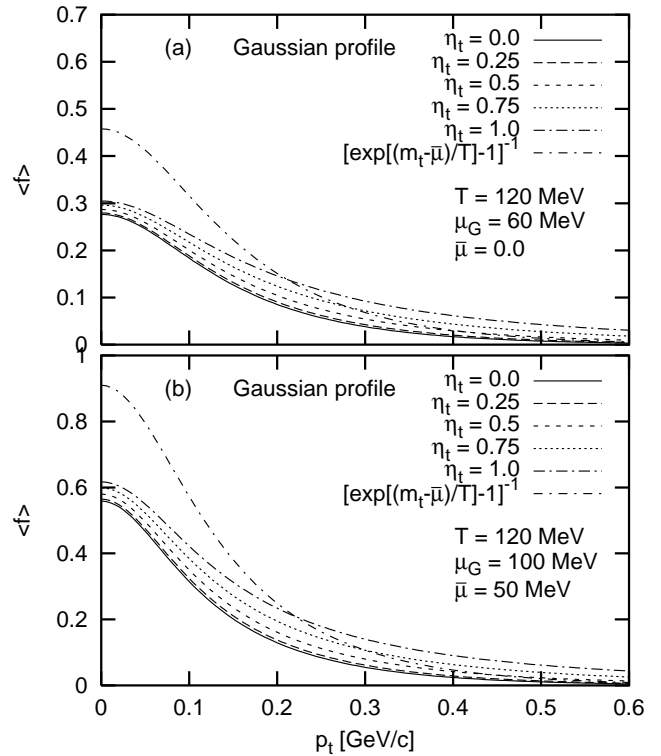,scale=1.7}
   \end {center}
\caption{Same as Fig.~\ref{F:flow} but for a Gaussian transverse 
density profile.
\label{F:Gauss}
}
}
\end{figure}

Figures \ref{F:flow} and \ref{F:Gauss} show that the longitudinal flow is 
again responsible for most of the suppression of $\fav$ compared to the 
local rest frame distribution. Transverse flow tends to flatten the 
$\pt$-dependence of $\langle f\rangle$ rather than to suppress $\langle
 f\rangle$. This effect stems from enhanced contributions at higher 
transverse momenta from outward-moving fluid cells; this is the 
well-known ``blueshift'' of the transverse momentum spectrum. The 
details of this flattening depend on the particular model for the 
transverse density profile.

The calculated $\fav$ in Figs.~\ref{F:flow} and \ref{F:Gauss} are 
compared to Bose-Einstein distributions with an {\em averaged} 
chemical potential $\bar \mu$ which takes into account the averaging 
over the transverse density profile. It is obtained by calculating the
``transverse average'' 
 \be{E:trav}
 \label{ftr}
   \bar f_{\rm tr}(\vp) = 
   \frac{\int_0^\infty r\,dr\, f_{\rm tr}^2(r,\vp)}%
   {\int_0^\infty r\,dr\, f_{\rm tr}(r,\vp)}
 \ee
of the underlying local distribution without flow 
 \be{E:trdist}
   f_{\rm tr}(r,\vp) = \frac{1}{\exp\left[ (m_\perp - \mu(r))/T \right ] -1}
   \, ,
 \ee 
where $\mu(r)$ characterizes the chosen transverse density distribution. 
Then a Bose-Einstein distribution
$f_{\rm BE} = 1/[\exp((m_\perp-\bar \mu)/T)-1]$ with the same temperature
is fitted to $\bar f_{\rm tr}(\vp)$, using $\bar \mu$ as a fit parameter to
ensure the same normalisation $\int f_{\rm BE}(m_\perp) m_\perp d_m\perp$."
Obviously, for the box profile $\bar\mu{\,=\,}\mu_B$; for 
the Gaussian profile, $\bar\mu{\,<\,}\mu_G$. For the Gaussian profile
the Bose fit is reasonable but not perfect since $\bar f_{\rm tr}(\vp)$ 
is a bit more strongly peaked at small $p_\perp$ than for the box profile. 
However, since $\bar\mu/T$ controls the normalization of $f$, i.e. the 
total pion yield at midrapidity $y{\,=\,}0$, $\bar\mu$ is identical with 
the {\em global} chemical potential extracted from a chemical analysis 
of midrapidity particle abundances at the {\em thermal} freeze-out 
temperature, and therefore a good reference value. 

For the direct comparison of the two transverse density profiles we 
proceed as follows: we fix $\bar \mu$ and, for the Gaussian profile, 
adjust $\mu_G$ accordingly. A comparison of the solid lines in panels 
(a,b) of Figs.~\ref{F:flow} and \ref{F:Gauss} shows that in the absence
of transverse flow, at the same value of $\bar\mu$, the dilution effects 
on $\fav$ from longitudinal expansion are almost identical for both
density profiles. However, the additional flattening effects from 
transverse flow are weaker in the Gaussian model than for the box profile. 

We have demonstrated that the thermal model of ultrarelativistic
heavy ion collisions does {\em not} predict an average phase-space 
density characterised by a Bose-Einstein distribution function. 
Comparing data for $\fav$ extracted from Eq.~(\ref{1}) with such a 
distribution is likely to severely underestimate the pion chemical 
potential or temperature. The previously used {\em ad hoc} introduction 
of a naive transverse blue-shift factor \cite{R:phd,R:cramer}, 
without first properly accounting for the dominant effects from 
longitudinal expansion, is based on faulty intuition and misleading. 
Longitudinal expansion strongly reduces the average freeze-out 
phase-space density at low $\pt$, whereas transverse flow leads to 
an additional flattening of $\la f\ra(\pt)$. At low $\pt$, where
$\fav$ and thus the danger of distortions due to strong multi-boson 
symmetrization effects \cite{R:hsz} are largest, the strong longitudinal 
expansion of the heavy-ion collision fireball dominates the suppression
of the average phase-space density. An extraction of the pion chemical 
potential should be based on a comparison of the measured values for 
$\fav$ with Eqs.~(\ref{E:results}), or a suitable generalisation thereof 
which more accurately accounts for shortlived resonance decays (which we 
treated here rather superficially).  

{\bf Acknowledgements:} We thank John Cramer and Jim Draper for stimulating 
discussions and valuable comments on a first draft of this paper. The 
research of U.H. is supported by the U.S. Department of Energy under contract 
DE-FG02-01ER41190.


\end{document}